# Behavioral System Level Power Consumption Modeling of Mobile Video Streaming Applications

Yahia Benmoussa*[+], Jalil Boukhobza[+], Yassine Hadjadj Aoul[x], Djamel Benazzouz*, Loïc Lagadec[+]

[+]Université Européenne de Bretagne
Université de Brest ;
CNRS; UMR 3192 Lab-STICC,
20 avenue Le Gorgeu
29285 Brest Cedex 3, France
firstname.lastname@univ-brest.fr

* Université M'hamed
Bougara de Boumerdes,
LMSS
Avenue du 1er Novembre.
Boumerdes. Algeria.
yahia.benmoussa@gmail.com,

[x]INRIA Campus Universitaire
de Beaulieu, 35042 Rennes
Cedex, France
Yassine.hadjadj-aoul@inria.fr

*Abstract*— Nowadays, the use of mobile applications and terminals faces fundamental challenges related to energy constraint. This is due to the limited battery lifetime as compared to the increasing hardware evolution. Video streaming is one of the most energy consuming applications in a mobile system because of its intensive use of bandwidth, memory and processing power. In this work, we aim to propose a methodology for building and validating a high level global power consumption model including a hardware and software elements. Our approach is based on exploiting the interactions between power consumption sub-models of standalone systems in the perspective to build more accurate global model. The interactions are studied within the exclusive context of video streaming applications that are one of the most used mobile applications.

## I. INTRODUCTION

Mobile devices such as smartphones and tablets are increasingly becoming the most important channel for delivering Internet traffic. One of the most popular use of these terminals is streaming video. According to [1], during the few next years, this application will represent more than 70% of the overall internet mobile traffic.

The transmission and decoding of large amounts of video data on mobile devices requires more and more CPU processing, memory and network bandwidth. For instance, nowadays, a smartphone equipped with 1 Giga-hertz processor, 1 Gigabyte of RAM and 32 GB of flash is a standard configuration.

Running mobile applications, especially streaming on such hardware is very energy-intensive. In addition, battery technologies are not progressing at the same rate as the power consumption of the hardware. Consequently, battery life time is decreasing drastically and makes the problematic of power optimization one of the fundamental challenges the designers of mobile devices are facing.

In this paper we address, from a high level modeling perspective, the problematic of modeling and optimizing a streaming system including hardware components (CPU, RAM, Flash, and WNIC) and software (Operating System and streaming application).

## II. POWER CONSUMPTION MODELING LEVEL

Power consumption can be modeled using different approaches according to the abstraction level we consider for describing the overall system. According to [2] power models can be classified within one of the following levels: Physic, RTL (Register Transfer Level), architectural or behavioral. The first three levels are hardware dependent and are generally used during hardware design process to improve its consumption proprieties. Behavioral level intervenes at run time step and consists of studying how the manner the hardware is used is impacting its power consumption. It is the most dynamic model and it can include both hardware and software components. It is consequently suitable to take into consideration all the interactions which may occur between them.

## III. TARGET SYSTEM SCOPE & MOTIVATIONS

Power consumption is a critical issue in mobile terminals and generally in any embedded system that uses a battery. Many works have proposed solutions by addressing separately each one of the hardware component of the overall system. We can cite for example [3] for processors, [4] for WNIC and [5] for RAM. Other studies have considered subsystems that consist of two or more system components rather than studying a standalone element. For example, in [6], they study power consumption model of the processor when executing MPEG Streaming.

In this work, we model a complete streaming system composed of a standard architecture of a mobile terminal including processor, RAM, Flash memory and WNIC in addition to the applicative layer including both streaming client and operating system (Android and embedded Linux). We have also decided to study, in our model, the video codecs and network protocols and their impact on the power consumption.

From modeling level perspective, we consider all aspects related to the behavior of the hardware combined to software components. According to ITRS roadmap [7], more than 50% optimization effort will be concentrated at this level during few next years.

## IV. TARGET SYSTEM MODELING APPROACH

In our modeling approach we define *local* models, *interaction* and *global* model entities. By local model we mean a model $M_{Li}$ developed for a standalone hardware component regardless of the overall system which it belongs to. A local model serves to understand the behavior of power consumption of a system element in term of a set of parameters $\alpha_{i1}, \alpha_{i2}, \dots \alpha_{iN}$ which verify the following conditions: 1- They must be high level (didn't belong to physic, RTL or architecture level), 2- accessible from application and/or operating system and 3- shared (not necessary all of them) with other local models. We define the last condition as *interaction* between models.

In fact, we believe that considering the interaction between the local models in the overall system modeling process will be more benefit from power saving perspective than considering each model separately. Our approach is illustrated in Figure 1. For example, coupling DVFS technique [3] of the CPU, streaming codec, bitrate and DPM (Dynamic Power Management) [2] of the WNIC may bring considerable power saving.

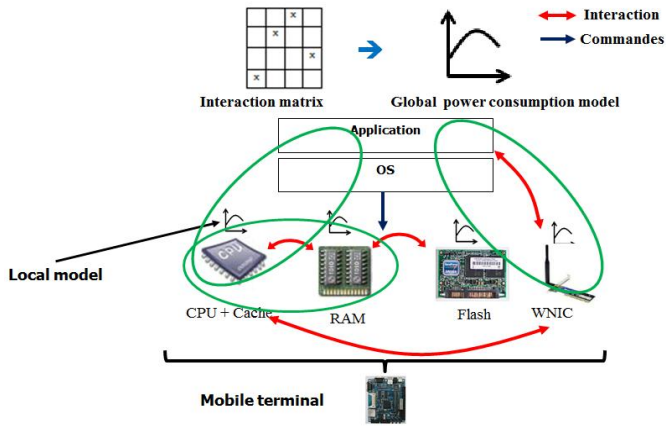

Figure11. Global modeling approach

## V. NOVEL STREAMING STANDARDS AND IMPACT ON ENERGY CONSUMPTION

The video streaming is at the center of our modeling approach because : 1- we consider the power consumption problematic in a context of exclusive use of video streaming application and 2- data streams across all hardware components and thus streaming parameters (bitrate, video quality, resolution) may impact local models of WNIC, CPU and memories components. Thus, Streaming is source of interactions.

Video streaming consist of transporting compressed video data over the network. The compression is done using video codec which are transported by means of network protocols.

In this work we explore the opportunity offered by novel streaming standards in terms of scalability [8,9] to realize more power optimization while maintaining video quality delivery.

Scalable streaming is a technique allowing to select/extract dynamically a subset of video data to adapt to changes in terminal resource capabilities (screen resolution, processing, and power) or network change in terms of bandwidth.

## VI. MODEL VALIDATION AND OPTIMIZATION METHODOLOGY

The validation of our model is done using a set of power measurement tools developed within OpenPeople national project [10]. The Validation methodology is a cycle process composed of three steps: 1- measuring power consumption of local models on which we have initial knowledge. 2- Validation of local models and interactions. 3- Tuning the *interactions matrix* and the global model.

The tuned model is tested through the implementation of deduced optimizations under the assumption of its correctness. We suppose that valid model leads to pertinent optimizations and thus a gain in term of total power consumption. Figure-2 shows the above methodology.

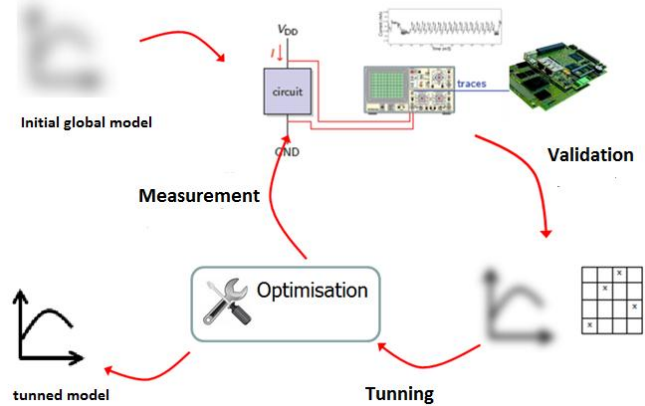

Figure-2. Model validation process of power consumption global model